\documentclass[12pt,a4paper]{article}
\usepackage[british]{babel}
\usepackage{epsfig}
\begin{document}
\date{}

\title{
{\vspace{-20mm} \normalsize
\hfill \parbox[t]{50mm}{\small DESY 12-194 }}        \\[25mm]
 Unquenched simulations of four-nucleon interactions \\[5mm]}
\author{ I.\ Montvay                             \\
         Deutsches Elektronen-Synchrotron DESY   \\
         Notkestr.\,85, D-22603 Hamburg, Germany }

\newcommand{\be}{\begin{equation}}                                              
\newcommand{\ee}{\end{equation}}                                                
\newcommand{\half}{\frac{1}{2}}                                                 
\newcommand{\rar}{\rightarrow}                                                  
\newcommand{\lar}{\leftarrow}
\newcommand{\LCB}{\raisebox{-0.3ex}{\mbox{\LARGE$\left\{\right.$}}}
\newcommand{\RCB}{\raisebox{-0.3ex}{\mbox{\LARGE$\left.\right\}$}}}
\newcommand{\U}{\mathrm{U}}
\newcommand{\SU}{\mathrm{SU}}
                                                                                
\maketitle
\vspace*{1em}

\begin{abstract} \normalsize
 Exploratory simulations of four-nucleon interactions are performed
 taking into account the dynamical effects of internal nucleon loops.
 The four-nucleon interactions in the isoscalar and isovector channels
 are described by Yukawa interactions with auxiliary scalar fields.
 The nucleon mass and the average field lengths of the scalar fields are
 determined as a function of nucleon hopping parameter and Yukawa
 coupling strengths.
 There are no problems with ``exceptional configurations'' at strong
 couplings which make quenched simulations unreliable.
\end{abstract}       

\section{Introduction}\label{sec1}
 There are impressive recent developments in solving QCD for the lightest
 nuclei, using the numerical technique of Lattice QCD (LQCD)
 (for recent reviews see \cite{SAVAGE,AOKI}).
 In the QCD-approach the nucleons are described as bound states of
 quarks and gluons and the nuclei are composed out of these composite
 objects.
 For nuclei heavier than, say, the helium an approach based on quarks and
 gluons would require a tremendous computing power beyond the
 exaflops-scale.
 However, in order to compute, for instance, binding energies of heavier
 nuclei it is also not clear whether one has to stick to the fundamental
 theory, because the relevant degrees of freedom are not quarks and
 gluons but nucleons, pions and other hadrons described in a chiral
 effective theory \cite{WEINBERG:1990,WEINBERG:1991}.

 The chiral effective theory can be taken as a starting point for
 applying non-perturbative lattice thechniques for nuclear physics.
 A lattice approach to nuclear physics was presented in a row of papers
 by the authors of \cite{Borasoy:2005yc}-\cite{Epelbaum:2011md}.
 For an overview over the various available approaches and methods to
 nuclear physics using chiral effective theories see the recent review
 articles \cite{Lee:2008fa}-\cite{Machleidt:2011zz}.
 The approach taken in these publications differs fundamentally from
 the methods applied in LQCD simulations where the path integral in
 Euclidean space-time is simulated using Monte Carlo methods.

 Recently, the application of Monte Carlo methods of LQCD has also been
 tried in \cite{MONTVAY-URBACH} (for an earlier attempt see also
 \cite{LEE-SCHAFER}).
 In Ref.~\cite{MONTVAY-URBACH} two-nucleon binding energies have been
 calculated in the so called {\it quenched approximation} to the
 pionless effective field theory with isoscalar and isovector auxiliary
 scalar fields.
 In this approximation the fermion determinant in the path integral
 is replaced by a constant, i.e. internal fermion loops are neclected
 and hence no dynamical effects of the fermions are taken into account.
 Apart from being only an approximation to the Euclidean path integral,
 the quenched approximation suffers from the so called {\it exceptional
 configurations} -- well known also from lattice QCD
 \cite{WUPPERTAL,Bardeen:1997gv}.
 Such field configurations have extremely small eigenvalues of the
 fermion matrix and hence give huge contributions to some expectation
 values and completely spoil the statistics.
 As is has been concluded in Ref.~\cite{MONTVAY-URBACH}, in the effective
 field theory for nuclear interactions the occurence of exceptional
 configurations make quenched simulations impossible for strong
 couplings (actually for bare couplings in the range
 $\max(|C_0|,|C_1|) \geq 0.3$).
 This phenomenon has also been observed earlier in quenched simulations
 of some simpler Yukawa-models in \cite{deSoto:2006jr}-\cite{deSoto:2011aa}.
 Since this problem does not appear in numerical simulations of Yukawa
 models with dynamical fermions \cite{FARAKOS-YUKAWA,LIN-YUKAWA},
 it can be expected that it does also disappear in nuclear Yukawa models
 if dynamical nucleons are included in the simulation update.

 In this paper the results of first unquenched Euclidean Monte Carlo
 simulations are presented in the model studied previously in
 \cite{MONTVAY-URBACH} with particular emphasis on the strong coupling
 region $\max(|C_0|,|C_1|) \geq 0.3$.
 After defining the lattice action and shortly describing the numerical
 simulation algorithm in Section~\ref{sec2}, the results are presented.
 Conclusions and the discussion of the outlook for future work are
 contained in Section~\ref{sec3}.

\section{Numerical simulations}\label{sec2}
\subsection{Lattice action and updating}\label{sec2.1}

 The four-nucleon interactions are described here, as in
 Ref.~\cite{MONTVAY-URBACH} by Yukawa-couplings of the nucleon field to
 auxiliary scalar fields with isospin 0 ($\phi^{(0)}_x$) and 1
 ($\phi^{(1)}_{ax}\;(a=1,2,3)$).
 The nucleon fields are described by a pair of Grassmann variables and are
 denoted by $\psi_{\alpha x}$ and $\tilde{\psi}_{\alpha x}$ where
 $\alpha=1,2$ is the isospin index.
 The lattice action is defined by
\be\label{eq2.1}
S = S_N + S_{NA} \,.
\ee
 where $S_N$ is the Wilson fermion lattice action \cite{WILSON}
 with the nucleon hopping parameter $\kappa$:
\be\label{eq2.2}
S_{N} = \sum_x \left\{ 
(\tilde{\psi}_x \psi_x) - \kappa \sum_{\mu=\pm 1}^{\mu=\pm 4}
(\tilde{\psi}_{x+\hat{\mu}} [1+\gamma_\mu] \psi_x) \right\} \,.
\ee
 and the scalar part of the action contains the Yukawa-type couplings
 to the nucleon field:
\be\label{eq2.3}
S_{NA} = \sum_x \left\{
\phi^{(0)}_x\phi^{(0)}_x + \phi^{(1)}_{a x}\phi^{(1)}_{a x} +
C_0\, \phi^{(0)}_x\, (\tilde{\psi}_x \psi_x) +
C_1\, \phi^{(1)}_{a x}\, (\tilde{\psi}_x \tau_a \psi_x) \right\} \,.
\ee
 Here $C_0$ and $C_1$ are the bare Yukawa-couplings,
 $\tau_a , (a=1,2,3)$ are Pauli-matrices for isospin and a summation
 over repeated indices $a$ is understood.
 The four-nucleon interactions are obtained after integrating over the
 auxiliary fields according to
\be\label{eq2.4}
\int_{-\infty}^\infty d \phi\, 
\exp\{-\phi^2 - C\, \phi (\tilde{\psi} \psi) \} =
\sqrt{\pi}\, \exp\{ \frac{C^2}{4}\, (\tilde{\psi} \psi)^2 \} \,.
\ee

 In Ref.~\cite{MONTVAY-URBACH} block fields were introduced in order to
 separate the physical cut-off due to the extended nature of hadrons
 from the inherent lattice cut-off given by the inverse of the lattice
 spacing $a$.
 In the present paper the block fields are not introduced and there is
 a unique lattice cut-off.
 The (straightforward) introduction of the block fields is left for future
 work.

 For the creation of sequences of interacting scalar fields 
 $\phi^{(0)}_x ,\; \phi^{(1)}_{ax}$ the {\it two-step Polynomial
 Hybrid Monte Carlo} TSPHMC \cite{MONTVAY:TSMB,MONTVAY-SCHOLZ,SCHOLZ-MONTVAY}
 algorithm is used which is a suitably adapted version of the original
 HMC \cite{HMC} and, in particular, PHMC \cite{PHMC} algorithms.

\subsection{Numerical simulation results}\label{sec2.2}

 The numerical simulations are performed on a $8^3 \cdot 32$ lattice.
 This kind of lattices can be simulated on PC's (in our case
 on the PC cluster of the DESY theory group).
 The number of trajectories in the simulation samples is at least
 5000, in some cases up to 15000.
 The nucleon mass $m_N$ and the average lengths squared of the scalar fields
\be\label{eq2.5}
\Phi_0 \equiv
\left\langle \frac{1}{N_x}  \sum_x \phi^{(0)}_x \phi^{(0)}_x \right\rangle
\, ,\hspace{3em}
\Phi_1 \equiv
\left\langle \frac{1}{N_x}  \sum_x \phi^{(1)}_{ax} \phi^{(1)}_{ax} \right\rangle
\, .
\ee
 are determined as functions of the three bare parameters $\kappa,C_0,C_1$.
 ($N_x$ denotes the number of lattice sites.)

\begin{figure}[t]
\begin{center}
{\Large\bf Figures}
\end{center}
\vspace*{1em}
\vspace*{0.01\vsize}
\begin{minipage}[c]{1.0\linewidth}
\begin{flushleft}
\hspace{-0.01\hsize}
\includegraphics[angle=-90,width=0.95\hsize]
 {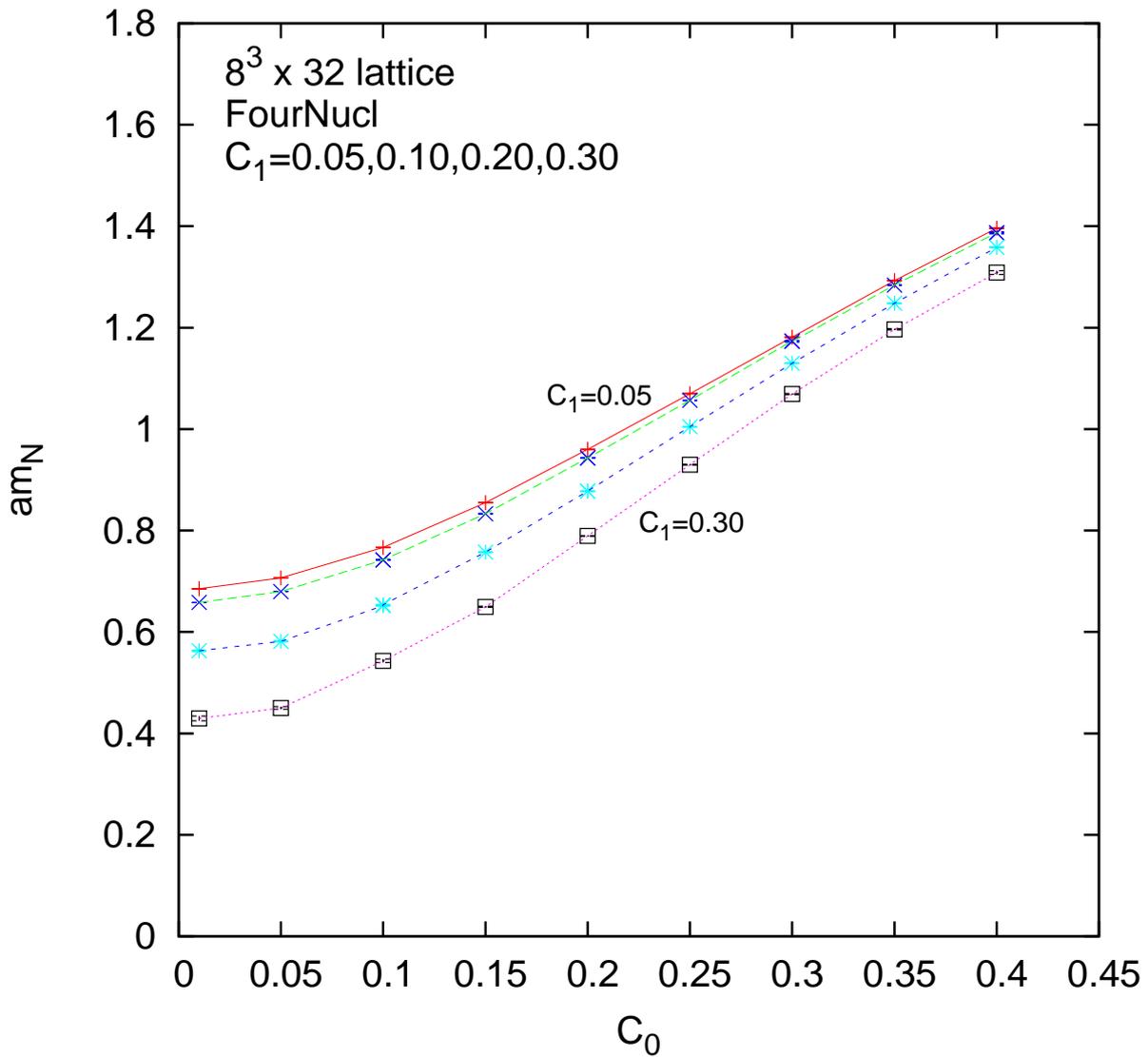}
\end{flushleft}
%
\end{minipage}
\vspace*{-0.1em}
\begin{center}
\parbox{0.8\linewidth}{\caption{\label{fig1}\em
 The nucleon mass as function of the coupling $C_0$
 for $\kappa=0.10$ and different values of $C_1$.
 The lines are drawn to guide the eyes.} }
\end{center}
\vspace*{-1em}
\end{figure}

\begin{figure}[t]
\vspace*{0.01\vsize}
\begin{minipage}[c]{1.0\linewidth}
\begin{flushleft}
\hspace{-0.01\hsize}
\includegraphics[angle=-90,width=0.95\hsize]
 {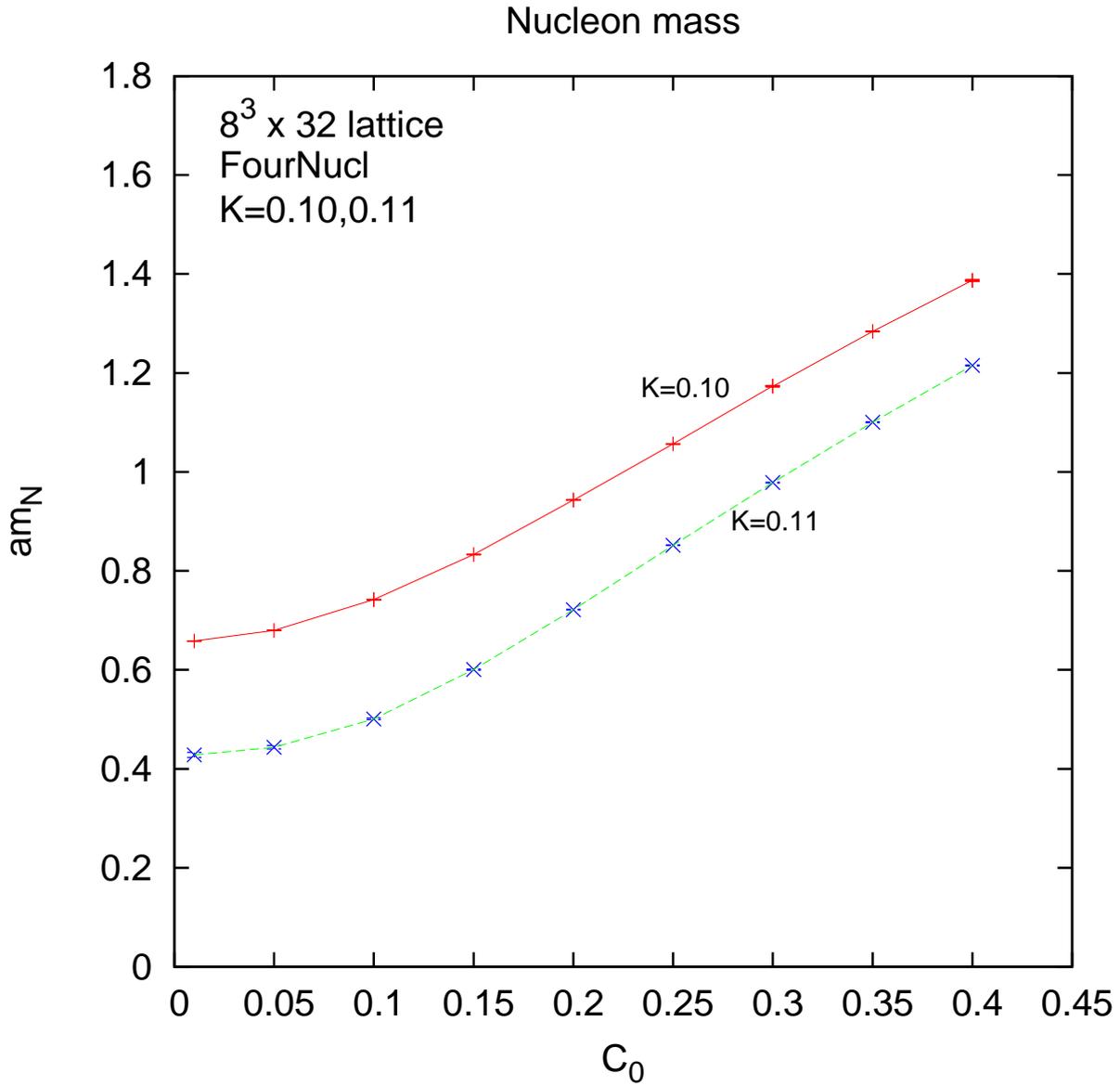}
\end{flushleft}
\end{minipage}
\vspace*{-0.1em}
\begin{center}
\parbox{0.8\linewidth}{\caption{\label{fig2}\em
 The nucleon mass as function of the coupling $C_0$
 for $\kappa=0.10$ and $\kappa=0.11$.
 The lines are drawn to guide the eyes.} }
\end{center}
\vspace*{-1em}
\end{figure}

\begin{figure}[t]
\vspace*{0.01\vsize}
\begin{minipage}[c]{1.0\linewidth}
\begin{flushleft}
\hspace{-0.01\hsize}
\includegraphics[angle=-90,width=0.95\hsize]
 {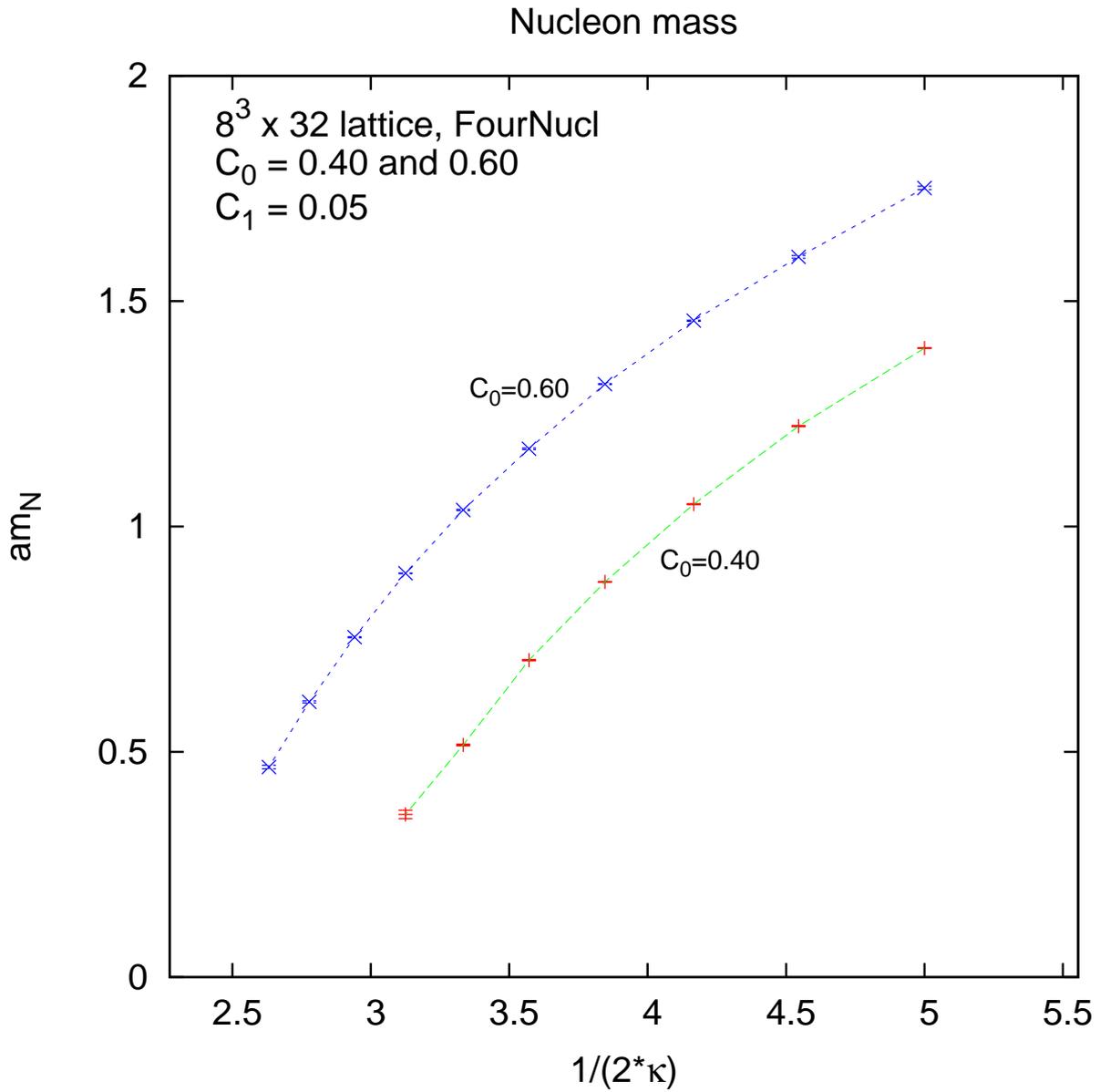}
\end{flushleft}
\end{minipage}
\vspace*{-0.1em}
\begin{center}
\parbox{0.8\linewidth}{\caption{\label{fig3}\em
 The nucleon mass as function of the bare mass parameter
 $(2\kappa)^{-1}$ for $C_1=0.05$ and for $C_0=0.40$ and $C_0=0.60$.
 The lines are drawn to guide the eyes.} }
\end{center}
\vspace*{-1em}
\end{figure}

\begin{figure}[t]
\vspace*{0.01\vsize}
\begin{minipage}[c]{1.0\linewidth}
\begin{flushleft}
\hspace{-0.01\hsize}
\includegraphics[angle=-90,width=0.95\hsize]
 {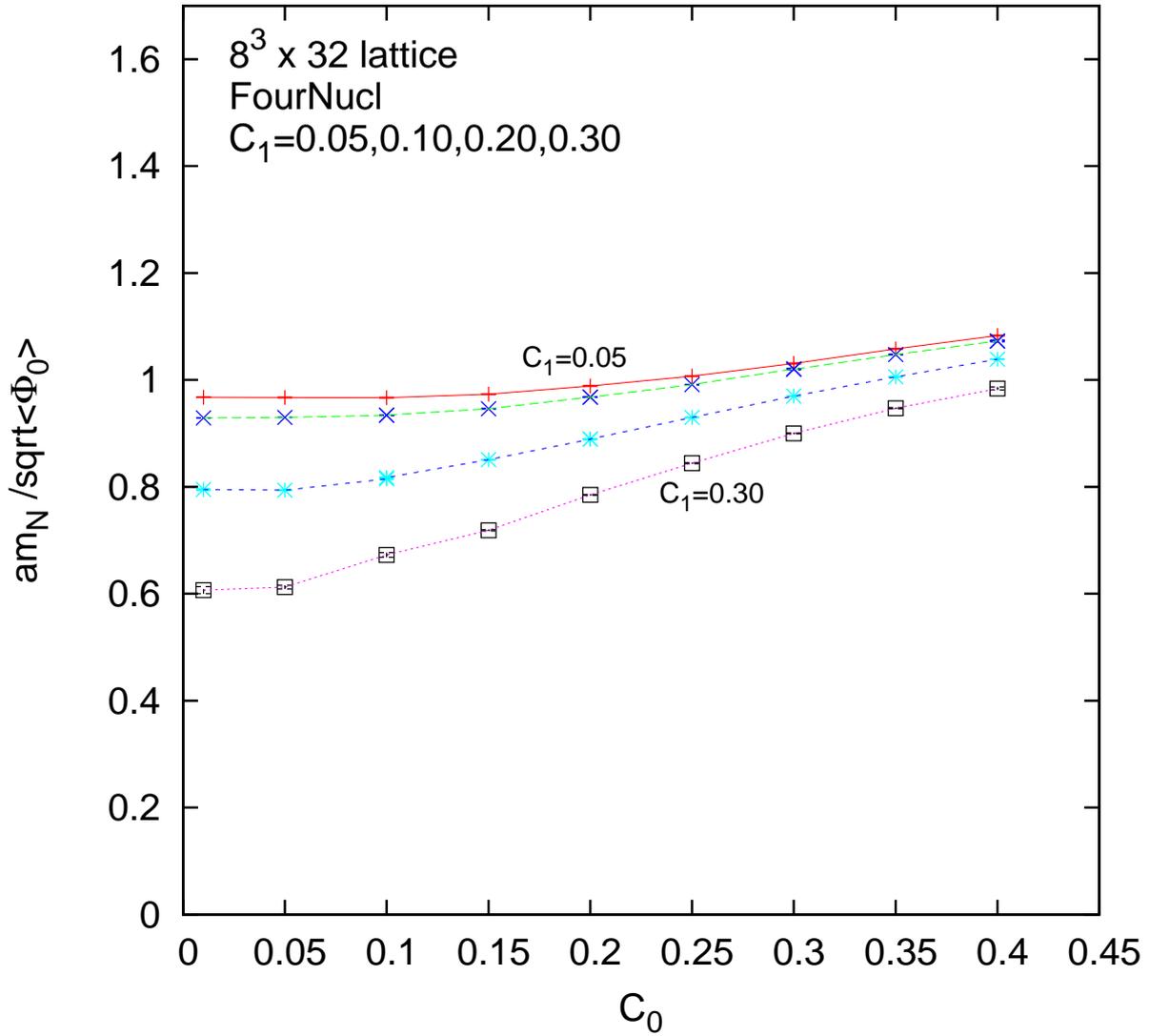}
\end{flushleft}
%
\end{minipage}
\vspace*{-0.1em}
\begin{center}
\parbox{0.8\linewidth}{\caption{\label{fig4}\em
 The nucleon mass divided by the expectation value of the average length
 of the isoscalar field $\langle \Phi_0 \rangle^{1/2}$ as function of
 the coupling $C_0$ for $\kappa=0.10$ and different values of $C_1$.
 The lines are drawn to guide the eyes.} }
\end{center}
\vspace*{-1em}
\end{figure}

\begin{figure}[t]
\vspace*{0.01\vsize}
\begin{minipage}[c]{1.0\linewidth}
\begin{flushleft}
\hspace{-0.01\hsize}
\includegraphics[angle=-90,width=0.95\hsize]
 {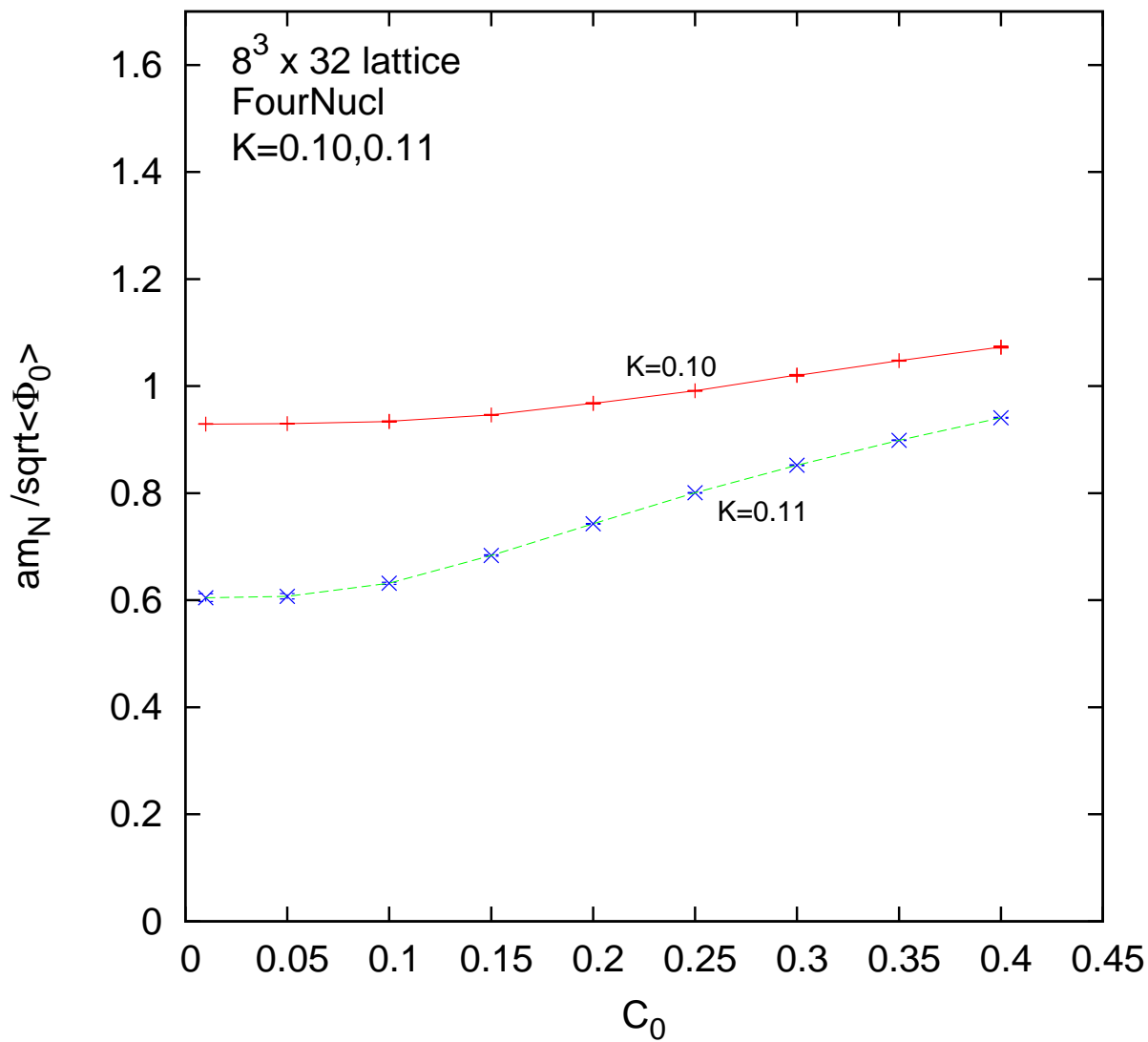}
\end{flushleft}
\end{minipage}
\vspace*{-0.1em}
\begin{center}
\parbox{0.8\linewidth}{\caption{\label{fig5}\em
 The nucleon mass divided by the expectation value of the average length
 of the isoscalar field $\langle \Phi_0 \rangle^{1/2}$ as function of
 the coupling $C_0$ for $\kappa=0.10$ and $\kappa=0.11$.
 The lines are drawn to guide the eyes.} }
\end{center}
\vspace*{-1em}
\end{figure}

\begin{figure}[t]
\vspace*{0.01\vsize}
\begin{minipage}[c]{1.0\linewidth}
\begin{flushleft}
\hspace{-0.01\hsize}
\includegraphics[angle=-90,width=0.95\hsize]
 {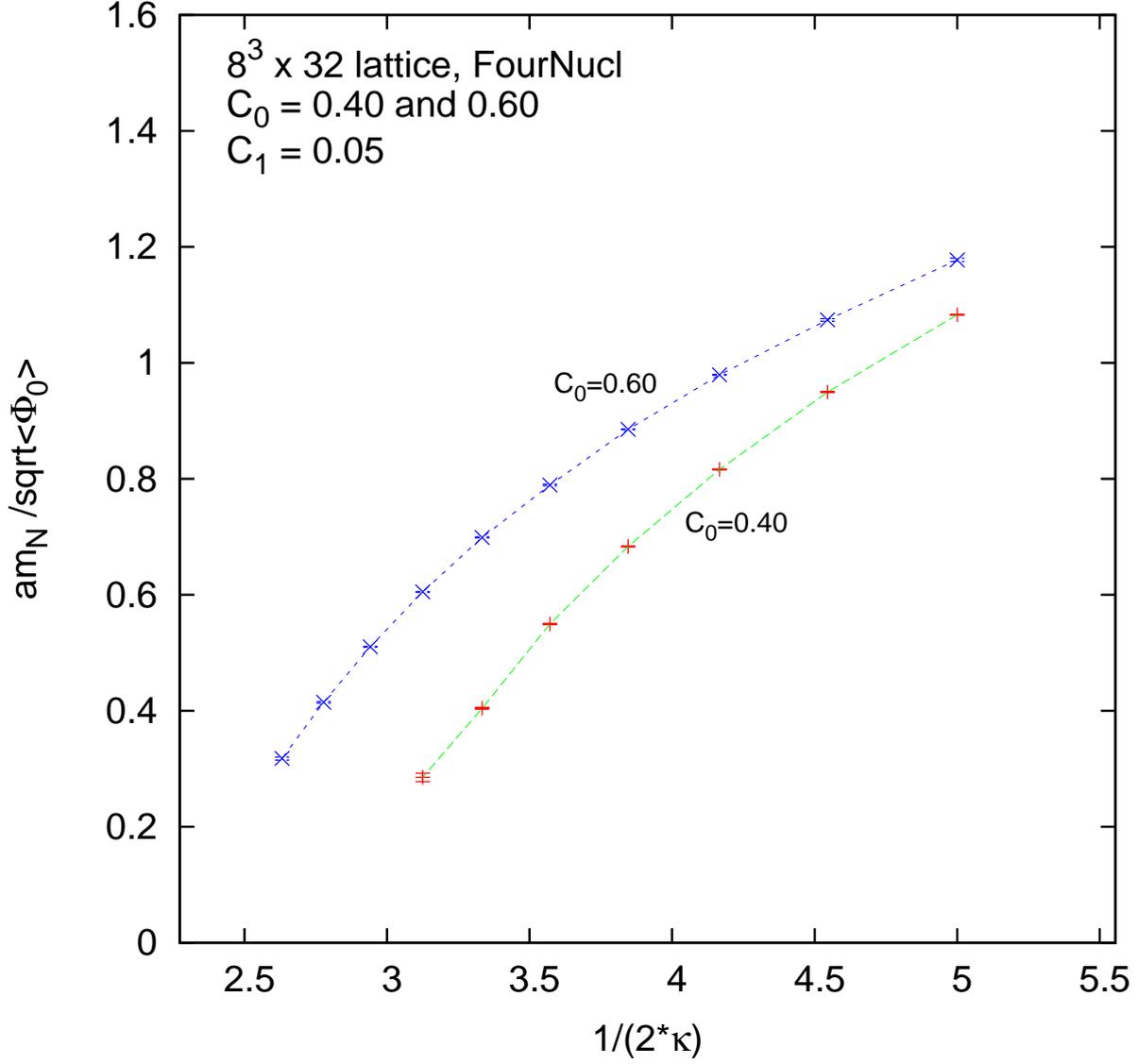}
\end{flushleft}
\end{minipage}
\vspace*{-0.1em}
\begin{center}
\parbox{0.8\linewidth}{\caption{\label{fig6}\em
 The nucleon mass divided by the expectation value of the average length
 of the isoscalar field $\langle \Phi_0 \rangle^{1/2}$ as function of the
 bare mass parameter $(2\kappa)^{-1}$ for $C_1=0.05$ and for $C_0=0.40$
 and $C_0=0.60$.
 The lines are drawn to guide the eyes.} }
\end{center}
\vspace*{-1em}
\end{figure}

 The results are summarized in Tables~\ref{tab1}-\ref{tab4} and illustrated
 in Figures~\ref{fig1}-\ref{fig6}.
 In general, the nucleon mass is increasing for increasing $C_0$
 and, somewhat surprisingly, decreasing for increasing $C_1$.
 The decrease for increasing hopping parameter is expected since
 $(2\kappa)^{-1}$ is, apart from an additive shift due to renormalization,
 the bare nucleon mass.
 This is well displayed by Fig.~\ref{fig3} which shows that for some
 {\em critical hopping parameter} $\kappa=\kappa(C_0,C_1)$ the nucleon
 mass becomes very small.
 (It is expected that in the infinite volume limit it becomes actually zero.)
 This is important if one wants to perform simulations in the small
 lattice spacing limit.

 It is an interesting question what kind of behaviour is observed if
 the nucleon mass is divided by the expectation value of the average length
 of the isoscalar field $\langle \Phi_0 \rangle^{1/2}$.
 As shown by Figs.~\ref{fig4}-\ref{fig5}, for small $C_0$ this ratio
 is approximately constant, but for larger values of $C_0$ there is
 a moderate increase.
 For fixed $C_0$ and $C_1$ as a function of the hopping parameter the average
 field length does not change much.
 As a consequance, the qualitative behaviour in Figs.~\ref{fig3} and \ref{fig6}
 is similar.

\section{Conclusion and outlook}\label{sec3}

 An important outcome of the performed unquenched simulations is
 the absence of any {\em exceptional configurations} with very small
 eigenvalues of the fermion matrix.
 No such configurations occured even at the strongest bare couplings
 which are much higher than the ones reached in Ref.~\cite{MONTVAY-URBACH}.
 The next goal should be to determine the renormalized four-nucleon
 couplings or, for instance, to tune the values of the dimensionless
 quantities $a_0 M_N$ and $\Delta/m_N$ where $a_0$ denotes the scattering
 length in the $^1S_0$ channel and $\Delta$ is the deuterium binding energy.
 This goes beyond the scope of this paper but is planned for future work.

 From the theoretical point of view it is an interesting question
 whether the tuning of the renormalized couplings to their physical values
 is possible at all by changing the values of the bare couplings $C_0$
 and $C_1$.
 The model studied in this paper is a special case of a renormalizable
 Higgs-Yukawa model where the kinetic terms of the scalar fields are
 omitted and the bare quartic self-couplings of the scalar fields are
 set to zero.
 These terms are all dynamically generated by the four-nucleon interactions.
 This means that the model investigated in this paper explores a subspace
 in the space of renormalized couplings of the complete renormalizable
 Higgs-Yukawa model.
 Higgs-Yukawa models in general are expected to have a trivial continuum
 limit with all renormalized couplings equal to zero.
 Therefore, in the complete Yukawa-model and also in  the subspace
 explored in this paper there is a lattice spacing (i.e. cut-off)
 dependent upper limit on the renomalized couplings which tend to
 zero if the lattice spacing goes to zero.
 (Zero lattice spacing means that the nucleon mass in lattice units
 $am_N$ is zero.)

 The model studied in this paper is, of course, incomplete and does
 not properly describe the nucleon interactions.
 First of all the pion field with its Yukawa interaction to the nucleons
 has to be introduced.
 Since the pion is even lighter than the nucleon, its kinetic term
 has to be included together with the quartic self interaction.
 For a complete description of the nuclear interactions further
 non-renormalized couplings have to be introduced which are generated
 within the framework of the chiral effective theory
 \cite{WEINBERG:1990,WEINBERG:1991}, which served as a basis of previous
 investigations of nuclear physics, for instance, in
 \cite{Borasoy:2005yc}-\cite{Machleidt:2011zz}.

 A further step towards a precise description of nuclear physics
 with Monte Carlo methods is to control lattice artifacts introduced
 by the discrete lattice which break Lorentz symmetry. 
 One possibility is to introduce higher dimensional (non-renormalizable)
 couplings to compensate for the lattice artifacts within the framework
 of {\em improved lattice actions}
 \cite{Symanzik:1981hc}-\cite{Symanzik:1983gh}.
 Of course, in this way a large number of couplings has to be introduced
 and properly tuned.
 Therefore, this approach becomes at some point rather  cumbersome.
 Another possibility is to introduce block fields \cite{MONTVAY-URBACH}
 which mimic the extended nature of hadrons (nucleons and pion) and
 introduce by their finite space-time extensions a natural cut-off in
 momentum space.
 In this way the physical cut-off can be separated from the lattice
 cut off and a small lattice spacing limit can be defined keeping
 the physical cut-off at its desired physical value.
 

\clearpage
\begin{table}
\begin{center}
{\Large\bf Tables}
\end{center}
\vspace*{1em}
\begin{center}
\parbox{0.8\linewidth}{\caption{\label{tab1}\em
 Numerical simulation results on $8^3 \cdot 32$ lattice for
 $\kappa=0.10$ and $C_1=0.05,0.10$.
 Statistical errors in last digits are given in parentheses.}}
\end{center}
\begin{center}
\renewcommand{\arraystretch}{1.2}
\begin{tabular}{*{7}{|l}|}
\hline
 \multicolumn{1}{|c|}{$\kappa$}    &
 \multicolumn{1}{|c|}{$C_0$}       &
 \multicolumn{1}{|c|}{$C_1$}       &
 \multicolumn{1}{|c|}{$\langle \Phi_0 \rangle$}  &
 \multicolumn{1}{|c|}{$\langle \Phi_1 \rangle$}  &
 \multicolumn{1}{|c|}{$am_N$}    \\
\hline\hline
 0.10 & 0.01 & 0.05  & 0.501442(86) & 1.48565(15) & 0.68519(10)  \\
\hline
 0.10 & 0.05 & 0.05  & 0.53377(10)  & 1.48538(16) & 0.70653(37)  \\
\hline
 0.10 & 0.10 & 0.05  & 0.62883(12)  & 1.48634(16) & 0.76671(40)  \\
\hline
 0.10 & 0.15 & 0.05  & 0.77137(15)  & 1.48714(16) & 0.85493(57)  \\
\hline
 0.10 & 0.20 & 0.05  & 0.94305(17)  & 1.48868(15) & 0.96010(72)  \\
\hline
 0.10 & 0.25 & 0.05  & 1.12759(19)  & 1.48929(15) & 1.06971(72)  \\
\hline
 0.10 & 0.30 & 0.05  & 1.31323(20)  & 1.49048(16) & 1.1813(10)   \\
\hline
 0.10 & 0.35 & 0.05  & 1.49272(22)  & 1.49126(15) & 1.2927(10)   \\
\hline
 0.10 & 0.40 & 0.05  & 1.66176(23)  & 1.49255(17) & 1.39602(80)  \\
\hline
 0.10 & 0.50 & 0.05  & 1.96256(27)  & 1.49353(17) & 1.5840(14)   \\
\hline\hline
 0.10 & 0.01 & 0.10  & 0.501517(84) & 1.44292(14) & 0.65804(19)  \\
\hline
 0.10 & 0.05 & 0.10  & 0.534201(93) & 1.44379(14) & 0.67974(28)  \\
\hline
 0.10 & 0.10 & 0.10  & 0.630929(42) & 1.446002(51)& 0.74205(18)  \\
\hline
 0.10 & 0.15 & 0.10  & 0.77541(16)  & 1.44939(16) & 0.83309(79)  \\
\hline
 0.10 & 0.20 & 0.10  & 0.949356(51) & 1.453549(45)& 0.94291(20)  \\
\hline
 0.10 & 0.25 & 0.10  & 1.13598(18)  & 1.45796(14) & 1.05644(79)  \\
\hline
 0.10 & 0.30 & 0.10  & 1.32251(21)  & 1.46198(16) & 1.17402(64)  \\
\hline
 0.10 & 0.35 & 0.10  & 1.50271(21)  & 1.46593(14) & 1.28402(95)  \\
\hline
 0.10 & 0.40 & 0.10  & 1.67156(24)  & 1.46914(17) & 1.3872(10)   \\
\hline
\end{tabular}
\end{center}
\end{table}

\begin{table}
\begin{center}
\parbox{0.8\linewidth}{\caption{\label{tab2}\em
 Numerical simulation results on $8^3 \cdot 32$ lattice for
 $\kappa=0.10$ and $C_1=0.20,0.30$.
 Statistical errors in last digits are given in parentheses.}}
\end{center}
\begin{center}
\renewcommand{\arraystretch}{1.2}
\begin{tabular}{*{7}{|l}|}
\hline
 \multicolumn{1}{|c|}{$\kappa$}  &
 \multicolumn{1}{|c|}{$C_0$}     &
 \multicolumn{1}{|c|}{$C_1$}     &
 \multicolumn{1}{|c|}{$\langle \Phi_0 \rangle$}  &
 \multicolumn{1}{|c|}{$\langle \Phi_1 \rangle$}  &
 \multicolumn{1}{|c|}{$am_N$}    \\
\hline\hline
 0.10 & 0.01 & 0.20  & 0.501340(96) & 1.28479(13)  & 0.5629(12)   \\
\hline
 0.10 & 0.05 & 0.20  & 0.536649(99) & 1.28722(16)  & 0.58162(80)  \\
\hline
 0.10 & 0.10 & 0.20  & 0.639301(76) & 1.296000(78) & 0.65369(54)  \\
\hline
 0.10 & 0.15 & 0.20  & 0.79218(16)  & 1.30891(13)  & 0.75732(85)  \\
\hline
 0.10 & 0.20 & 0.20  & 0.97365(12)  & 1.32375(10)  & 0.87765(60)  \\
\hline
 0.10 & 0.25 & 0.20  & 1.16692(22)  & 1.33980(14)  & 1.00478(60)  \\
\hline
 0.10 & 0.30 & 0.20  & 1.35767(24)  & 1.35509(15)  & 1.12985(85)  \\
\hline
 0.10 & 0.35 & 0.20  & 1.53994(24)  & 1.36975(15)  & 1.24829(94)  \\
\hline
 0.10 & 0.40 & 0.20  & 1.70999(27)  & 1.38275(17)  & 1.35859(87)  \\
\hline\hline
 0.10 & 0.01 & 0.30  & 0.501390(95) & 1.06681(11)  & 0.4296(47)  \\
\hline
 0.10 & 0.05 & 0.30  & 0.53978(12)  & 1.07226(11)  & 0.4501(25)  \\
\hline
 0.10 & 0.10 & 0.30  & 0.65177(14)  & 1.08780(11)  & 0.5431(36)  \\
\hline
 0.10 & 0.15 & 0.30  & 0.81674(18)  & 1.11161(12)  & 0.6495(19)  \\
\hline
 0.10 & 0.20 & 0.30  & 1.01076(22)  & 1.14042(14)  & 0.78928(71) \\
\hline
 0.10 & 0.25 & 0.30  & 1.21305(26)  & 1.17062(14)  & 0.92980(98) \\
\hline
 0.10 & 0.30 & 0.30  & 1.41081(29)  & 1.20080(16)  & 1.0690(10)  \\
\hline
 0.10 & 0.35 & 0.30  & 1.59643(29)  & 1.22882(16)  & 1.1968(13)  \\
\hline
 0.10 & 0.40 & 0.30  & 1.76870(31)  & 1.25445(18)  & 1.3087(37)  \\
\hline
\end{tabular}
\end{center}
\end{table}

\begin{table}
\begin{center}
\parbox{0.8\linewidth}{\caption{\label{tab3}\em
 Numerical simulation results on $8^3 \cdot 32$ lattice for
 $\kappa=0.11$ and $C_1=0.10$.
 Statistical errors in last digits are given in parentheses.}}
\end{center}
\begin{center}
\renewcommand{\arraystretch}{1.2}
\begin{tabular}{*{7}{|l}|}
\hline
 \multicolumn{1}{|c|}{$\kappa$}  &
 \multicolumn{1}{|c|}{$C_0$}     &
 \multicolumn{1}{|c|}{$C_1$}     &
 \multicolumn{1}{|c|}{$\langle \Phi_0 \rangle$}  &
 \multicolumn{1}{|c|}{$\langle \Phi_1 \rangle$}  &
 \multicolumn{1}{|c|}{$am_N$}    \\
\hline\hline
 0.11 & 0.01 & 0.10  & 0.501270(83) & 1.44445(14)  & 0.4281(50)  \\
\hline
 0.11 & 0.05 & 0.10  & 0.533467(94) & 1.44488(14)  & 0.4436(36)  \\
\hline
 0.11 & 0.10 & 0.10  & 0.62864(13)  & 1.44667(15)  & 0.5009(15)  \\
\hline
 0.11 & 0.15 & 0.10  & 0.77138(15)  & 1.44977(15)  & 0.6005(12)  \\
\hline
 0.11 & 0.20 & 0.10  & 0.94432(16)  & 1.45419(15)  & 0.72170(73) \\
\hline
 0.11 & 0.25 & 0.10  & 1.13103(15)  & 1.45808(13)  & 0.85157(58) \\
\hline
 0.11 & 0.30 & 0.10  & 1.31808(22)  & 1.46208(16)  & 0.97838(74) \\
\hline
 0.11 & 0.35 & 0.10  & 1.49865(21)  & 1.46600(15)  & 1.10036(79) \\
\hline
 0.11 & 0.40 & 0.10  & 1.66819(24)  & 1.46932(16)  & 1.21527(81) \\
\hline
\end{tabular}
\end{center}
\end{table}

\begin{table}
\begin{center}
\parbox{0.8\linewidth}{\caption{\label{tab4}\em
 Numerical simulation results on $8^3 \cdot 32$ lattice for
 $C_1=0.05$ and $C_0=0.40,0.60$.
 Statistical errors in last digits are given in parentheses.}}
\end{center}
\begin{center}
\renewcommand{\arraystretch}{1.2}
\begin{tabular}{*{7}{|l}|}
\hline
 \multicolumn{1}{|c|}{$\kappa$}  &
 \multicolumn{1}{|c|}{$C_0$}     &
 \multicolumn{1}{|c|}{$C_1$}     &
 \multicolumn{1}{|c|}{$\langle \Phi_0 \rangle$}  &
 \multicolumn{1}{|c|}{$\langle \Phi_1 \rangle$}  &
 \multicolumn{1}{|c|}{$am_N$}    \\
\hline\hline
 0.10 & 0.40 & 0.05  & 1.66176(23)  & 1.49255(17)  & 1.39402(80) \\
\hline
 0.11 & 0.40 & 0.05  & 1.65810(23)  & 1.49209(17)  & 1.2230(13)  \\
\hline
 0.12 & 0.40 & 0.05  & 1.65354(24)  & 1.49221(17)  & 1.04973(95) \\
\hline
 0.13 & 0.40 & 0.05  & 1.64713(23)  & 1.49224(17)  & 0.87697(84) \\
\hline
 0.14 & 0.40 & 0.05  & 1.63763(23)  & 1.49233(17)  & 0.7033(13)  \\
\hline
 0.15 & 0.40 & 0.05  & 1.62421(25)  & 1.49253(17)  & 0.5154(20)  \\
\hline
 0.16 & 0.40 & 0.05  & 1.60449(25)  & 1.49267(18)  & 0.3612(94)  \\
\hline\hline
 0.10 & 0.60 & 0.05  & 2.2115(24)   & 1.49573(13)  & 1.7512(36)  \\
\hline
 0.11 & 0.60 & 0.05  & 2.21459(29)  & 1.49453(14)  & 1.5983(31)  \\
\hline
 0.12 & 0.60 & 0.05  & 2.21269(31)  & 1.49462(18)  & 1.4567(11)  \\
\hline
 0.13 & 0.60 & 0.05  & 2.20967(33)  & 1.49470(17)  & 1.3159(11)  \\
\hline
 0.14 & 0.60 & 0.05  & 2.20585(32)  & 1.49470(18)  & 1.1725(19)  \\
\hline
 0.15 & 0.60 & 0.05  & 2.20051(32)  & 1.49471(18)  & 1.0368(14)  \\
\hline
 0.16 & 0.60 & 0.05  & 2.19391(27)  & 1.49474(19)  & 0.89612(85) \\
\hline
 0.17 & 0.60 & 0.05  & 2.18459(29)  & 1.49474(19)  & 0.7542(11)  \\
\hline
 0.18 & 0.60 & 0.05  & 2.17155(28)  & 1.49500(19)  & 0.6109(21)  \\
\hline
 0.19 & 0.60 & 0.05  & 2.15482(30)  & 1.49495(19)  & 0.4664(40)  \\
\hline
\end{tabular}
\end{center}
\end{table}

\clearpage

\end{document}